\documentclass[journal,twoside,web]{ieeecolor}
\usepackage{generic}
\usepackage{cite}
\usepackage{amsmath,amssymb,amsfonts}
\usepackage{algorithmic}
\usepackage{graphicx}
\usepackage{textcomp}
\usepackage{hyperref}

\begin{document}
\title{A Demographic-Conditioned Variational Autoencoder for fMRI Distribution Sampling and Removal of Confounds}
\author{Anton Orlichenko, Gang Qu, Ziyu Zhou, Anqi Liu, Hong-Wen Deng, Zhengming Ding, Julia M. Stephen, Tony W. Wilson, Vince D. Calhoun, \IEEEmembership{Fellow, IEEE},  and Yu-Ping Wang, \IEEEmembership{Senior Member, IEEE}
\thanks{Manuscript received on May 15, 2024. This work was supported in part by NIH grants (P20GM109068, P20GM144641, R01MH121101, R01MH104680, R01MH107354, R01MH103220, R01EB020407, R56MH124925, 5U19AG055373) and NSF grant (\#1539067). \textit{(Corresponding author: Yu-Ping Wang.)}}
\thanks{Anton Orlichenko, Gang Qu, and Yu-Ping Wang are with the Department of Biomedical Engineering, Tulane University, New Orleans, LA 70118. (e-mail: aorlichenko@tulane.edu, gqu1@tulane.edu, wyp@tulane.edu).}
\thanks{Ziyu Zhou and Zhengming Ding are with the Department of Computer Science, Tulane University, New Orleans, LA 70118. (e-mail: zzhou11@tulane.edu, zding1@tulane.edu).}
\thanks{Anqi Liu and Hong-Wen Deng are with the Center for Biomedical Informatics and Genomics, Tulane Integrated Institute of Data \& Health Sciences, Tulane University, New Orleans, LA 70112. (e-mail: aliu10@tulane.edu, hdeng2@tulane.edu).}
\thanks{Tony W. Wilson is with the Institute for Human Neuroscience, Boys Town National Research Hospital, Boys Town, NE. (email: tony.wilson@boystown.org).}
\thanks{Julia M. Stephen is with Mind Research Network, Albuquerque, NM. (email: jstephen@mrn.org).}
\thanks{Vince D. Calhoun is with the Tri-Institutional Center for Translational Research in Neuroimaging and Data Science (TReNDS), Georgia State University, Georgia Institute of Technology, Emory University, Atlanta, GA. (email: vcalhoun@gsu.edu).}
}

\maketitle

\begin{abstract}
\textit{Objective:} fMRI and derived measures such as functional connectivity (FC) have been used to predict brain age, general fluid intelligence, psychiatric disease status, and preclinical neurodegenerative disease. However, it is not always clear that all demographic confounds, such as age, sex, and race, have been removed from fMRI data. Additionally, many fMRI datasets are restricted to authorized researchers, making dissemination of these valuable data sources challenging. \textit{Methods:} We create a variational autoencoder (VAE)-based model, DemoVAE, to decorrelate fMRI features from demographics and generate high-quality synthetic fMRI data based on user-supplied demographics. We train and validate our model using two large, widely used datasets, the Philadelphia Neurodevelopmental Cohort (PNC) and Bipolar and Schizophrenia Network for Intermediate Phenotypes (BSNIP). \textit{Results:} We find that DemoVAE recapitulates group differences in fMRI data while capturing the full breadth of individual variations. Significantly, we also find that most clinical and computerized battery fields that are correlated with fMRI data are not correlated with DemoVAE latents. An exception are several fields related to schizophrenia medication and symptom severity. \textit{Conclusion:} Our model generates fMRI data that captures the full distribution of FC better than traditional VAE or GAN models. We also find that most prediction using fMRI data is dependent on correlation with, and prediction of, demographics. \textit{Significance:} Our DemoVAE model allows for generation of high quality synthetic data conditioned on subject demographics as well as the removal of the confounding effects of demographics. We identify that FC-based prediction tasks are highly influenced by demographic confounds.
\end{abstract}


\section{Introduction}
\label{sec:introduction}
\IEEEPARstart{f}{MRI} measures the time-varying blood oxygen level-dependent (BOLD) signal in the brain in order to infer coarse-grained neuronal activation \cite{Belliveau1991FunctionalMO}. It has traditionally been used to localize specific functions such as vision \cite{Cox2003-tu}, emotion\cite{Phan2002-nb}\cite{6789831}\cite{Koelsch2006-ci}, attention \cite{Coull1998-op}\cite{Pugh1996-tm}, and language \cite{Hernandez2001-ec} to discrete cortical areas. Functional connectivity (FC) is the temporal Pearson correlation of BOLD signal between different regions in the brain \cite{Van_den_Heuvel2010-ig}, and has been used to predict demographics such as age \cite{10002422}, sex\cite{ICER2020105444}\cite{9146335}, and race \cite{Orlichenko2023-ue}, as well as clinical assessments for schizophrenia diagnosis\cite{Wang2018-iv}\cite{Rashid2016-ct} and pre-clinical neurodegenerative disease \cite{MILLAR2022119228}. Many groups have also attempted to use FC as a biomarker to predict scholastic achievement or general fluid intelligence \cite{Qu2021EnsembleMR}. Besides Perason correlation, several other metrics have been used for the calculation of FC, including partial correlation \cite{Wang2016-yi} and distance correlation \cite{Hu2019-ky}. Others have experimented with calculating dynamic FC \cite{Rashid2016-ct}. With increasing clinical field strength, prediction based on FC or other fMRI metrics is poised to become increasingly important in the research and clinical settings \cite{9721204}.

Concurrently, generative models such as the DALL-E \cite{pmlr-v139-ramesh21a} and GPT \cite{Radford2019LanguageMA} series have created textual and visual content that in many cases is indistinguishable from human-generated work \cite{Mei2024-nv}. In the computer vision domain, generative models include generative adversarial networks (GANs) \cite{goodfellow2014generative}, variational autoencoders (VAEs) \cite{Kingma2014}, and most recently, diffusion-based models \cite{rombach2022high}. All of these methodologies have been applied to fMRI data for the improvement of predictive ability or for performing image to image translation \cite{Laino2022-gy}\cite{Kim2021-tn}. However, previous work has mostly overlooked subject demographics as potential input in their generative models. Since fMRI data access is often restricted to national data repositories by qualified researchers, we believe that generative models that can produce synthetic data are useful to more easily disseminate information, but only if they can re-create the demographic distribution and individual variation found in the fMRI datasets.
%This makes the data unavailable to many students and junior researchers not established in the field. 

It is well known that fMRI can be used to predict demographics such as age, sex, and race \cite{10002422} \cite{ICER2020105444} \cite{Orlichenko2023-ue}. It is also known that it is crucial to control for demographic confounds when performing statistical analysis \cite{Pourhoseingholi2012-sk} \cite{Skelly2012-dm}. In fact, many simpler models have provisions for regressing out confounds \cite{Yu2018-zq}. There is a question, however, as to whether fMRI-based prediction using more complicated models is solely due to demographic signal present in fMRI \cite{Orlichenko2023-ue}. To this end, we present a new generative model based on a VAE that decorrelates latent features from subject demographics (DemoVAE). It accomplishes this by forcing such correlations to be zero during training and injecting demographic information in the decoder after calculation of the latent features. We add classifier and regression-guided loss functions \cite{dhariwal2021diffusion} to ensure that synthetic samples contain demographics-associated features that are compatible with models trained on real data. We believe our model serves two purposes: 1) generation of representative synthetic data based on datasets that are not accessible to the general public, and 2) creation of fMRI latent features which are free from the confounding effects of demographics. It is also possible that DemoVAE can aid in data harmonization by removing site-specific effects \cite{Yu2018-he} through treating site location as a demographic. These capabilities are validated on two large datasets accessible to qualified researchers.

The rest of this manuscript is organized as follows. Section~\ref{sec:methods} gives a recapitulation of the theory of the DemoVAE model and our specific training methodology, as well as a description of the datasets and experiments performed. Section~\ref{sec:results} provides experimental results. Section~\ref{sec:discussion} discusses significant conclusions drawn from fMRI group differences found via DemoVAE and how they relate to existing work. Section~\ref{sec:conclusion} concludes with a summary of the work. We make the code publicly available at the link in the footnote \footnote{\url{https://github.com/aorliche/demo-vae/}}. An online demo is also available \footnote{\url{https://aorliche.github.io/DemoVAE/}}.

\begin{figure*} 
    \centering
    \includegraphics[width=\textwidth]{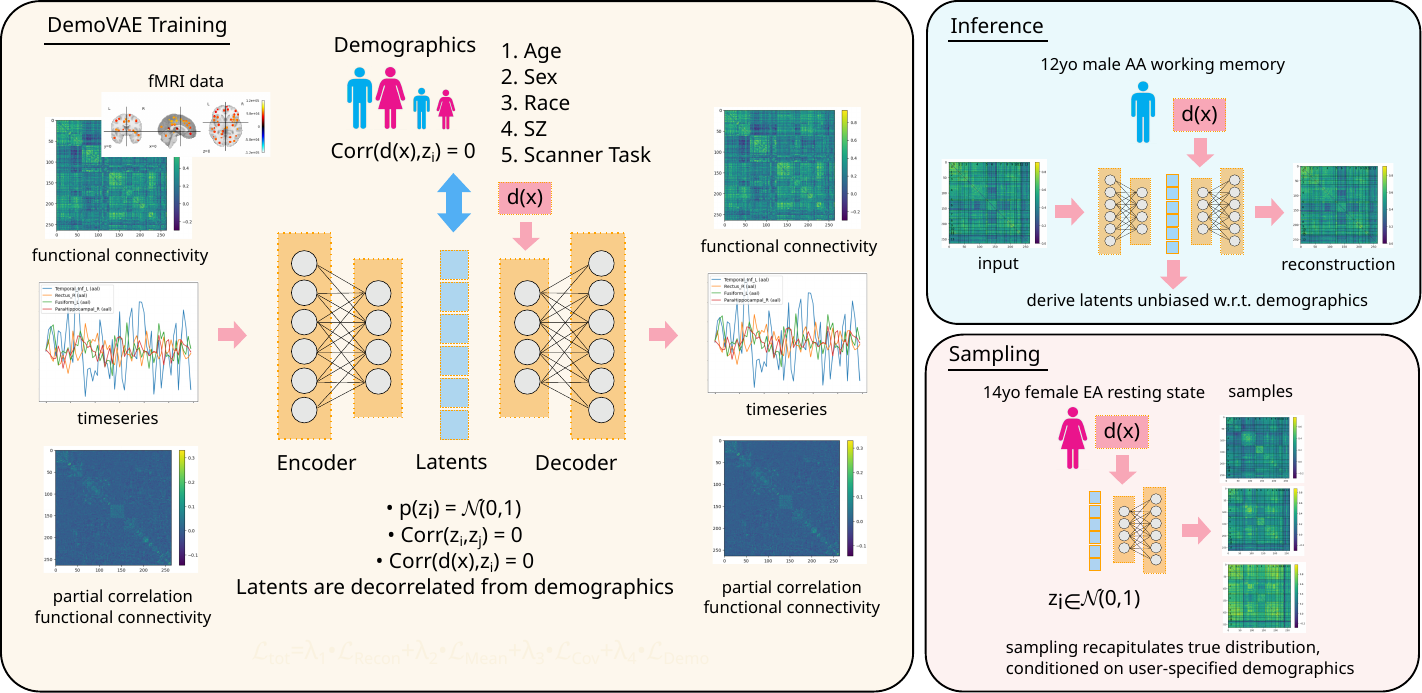}
    \caption{
    \label{fig:overview}Overview of the demographics-conditioned and decorrelated variational autoencoder (DemoVAE) model. Instead of reconstruction based only on latent features $\mathbf{z}=E_\phi(\mathbf{x})$, the DemoVAE model uses demographics $\mathbf{y}$ as input to the decoder $\hat{\mathbf{x}}=D_\theta(\mathbf{z},\mathbf{y})$. The two main uses of the model are inference, which generates latent features $\mathbf{z}$ decorrelated from demographics, and sampling, which generates synthetic fMRI data conditioned on user-provided demographics.}
\end{figure*}

\section{Methods}
\label{sec:methods}

First, we discuss the architecture and training of the DemoVAE model, shown in Figure~\ref{fig:overview}. Next, we describe two datasets used for the validation of the model. Then, we outline experiments used to analyze DemoVAE's ability to decorrelate latent features from demographic confounds as well as to generate high quality synthetic fMRI data. Finally, we describe experiments using DemoVAE for imputation of fMRI data.

\subsection{Variational Autoencoder}
\label{subsec:demovae}

An autoencoder (AE) converts raw features into a lower-dimensional latent space via a learned encoder function $\mathbf{z}=E_\phi(\mathbf{x})$, along with a decoder function to convert the latent features back into a reconstructed version of the input $\hat{\mathbf{x}}=D_\theta(\mathbf{z})$. The AE is often trained to minimize the difference between the reconstruction $\hat{\mathbf{x}}$ and original input $\mathbf{x}$. Thus, the AE may be seen as a nonlinear version of dimensionality reduction techniques such as PCA or Factor Analysis.

By contrast, a variational autoencoder (VAE) trains the encoder function $E_\phi(\mathbf{x})$ to produce latent features that approximate a known probability distribution $p_\theta(\mathbf{z})$, most often taken to be a standard multivariate Gaussian distribution $p_\theta(\mathbf{z})=\mathcal{N}(\mathbf{0},\mathbf{I})$ \cite{Kingma2014}. This allows for artificially constructing latent samples $\mathbf{z}_\text{samp}$ from the approximated distribution, followed by the conversion of those latents to samples of the original distribution $p_\theta(\mathbf{x}|\mathbf{z})$ by passing through the decoder function $\mathbf{x}_\text{samp}=D_\theta(\mathbf{z}_\text{samp})$.

For the following, consider scalar features $x$ and scalar latent features $z$. The exact calculation of $p_\theta(z|x)$ is in most cases intractable, therefore an approximation $q_\phi(z|x)\approx p_\theta(z|x)$ is made, and the Kullback-Leibler (KL) divergence between the two distributions is taken:

\begin{equation}
    \begin{split}
        D_{KL}(q_\phi(z|x)&\lVert p_\theta(z|x)) \\
        &= \mathbb{E}_{z\sim q_\phi(\cdot|x)}\left [\text{ln}\frac{q_\phi(z|x)}{p_\theta(z|x)}\right] \\
        &= \text{ln}p_\theta(x)+\mathbb{E}_{z\sim q_\phi(\cdot|x)}\left[\text{ln}\frac{q_\phi(z|x)}{p_\theta(x,z)}\right]
    \end{split}
\end{equation}

The evidence lower bound (ELBO) \cite{Kingma2014} is then defined as:

\begin{equation}
    \begin{split}
        L_{\theta,\phi} &= \mathbb{E}_{z\sim q_\phi(\cdot|x)}\left[\text{ln}\frac{p_\theta(x,z)}{q_\phi(z|x)}\right] \\
        &= \mathbb{E}_{z\sim q_\phi(\cdot|x)}\left[\text{ln}p_\theta(x|z) - D_{KL}(q_\phi(z|x)\lVert p_\theta(z))\right]
    \end{split}
\end{equation}

Maximizing the ELBO is equivalent to maximizing the reconstruction probability $\text{ln}p_\theta(x|z)$ while minimizing the KL divergence between our empirical and target distributions. Given a standard normal distribution for the latent features $p_\theta(z)=\mathcal{N}(0,1)$, the ELBO objective to be minimized becomes:

\begin{equation}
    \label{eq:elbo_loss}
    \mathcal{L}_{\theta,\phi} = \lVert x-D_\theta(z)\rVert^2_2 +
    N\sigma_z^2 + \lVert \mu_z\rVert^2_2 - N\text{ln}\sigma_z^2,
\end{equation}

\noindent where $\mu_z$ represents the mean of the the empirically calculated latent features, $\sigma_z^2$ represents the variance of the same, and $N$ represents the number of samples. This loss function can be seen to have three components: a reconstruction loss, two terms that tend to make $\sigma_z$ equal to one, and one term to make the expectation of the latent features equal to zero. Given this loss function, one is able to train a network to sample the distribution of FC data, but not to condition the samples on ancillary subject information such as demographics.

When considering a multivariate standard normal distribution $p_\theta(\mathbf{z})=\mathcal{N}(\mathbf{0},\mathbf{I})$ for the latent features, the KL divergence part takes the more complicated form \cite{pml2Book}:

\begin{equation}
\begin{split}
    D_{KL}&(\mathcal{N}(\mathbf{\mu}_z,\mathbf{\Sigma}_z)\lVert\mathcal{N}(\mathbf{0},\mathbf{I})) = \\
    & \frac{1}{2}\left [ \text{tr}(\mathbf{\Sigma}_z) + \mathbf{\mu}_z^\top\mathbf{\mu}_z + \text{log}(\text{det}( \mathbf{\Sigma}_z )) \right]
    \end{split}
\end{equation}

This presents a challenge due to the calculation of, and backpropagation through, the log determinant of the empirical latent covariance matrix $\mathbf{\Sigma}_z$. We address this issue as part of our modifications to the VAE loss function presented in Section~\ref{subsubsec:demovae}.

\subsection{Demographics-Conditioned and Decorrelated Variational Autoencoder (DemoVAE)}
\label{subsubsec:demovae}

There is an existing body of academic literature \cite{NIPS2015_8d55a249}\cite{DBLP:conf/nips/RazaviOV19} as well as practical applications \cite{ramesh2021zeroshot} exploring the conditioning of VAEs on user-specified inputs. VAEs have also been applied to the generation of synthetic fMRI data\cite{Kim2021-tn}, but without considering patient demographics. In this work, we include the known patient demographic features as input to the decoder function $\hat{\mathbf{x}}=D_\theta(\mathbf{z},\mathbf{y})$, where $\mathbf{z}$ are the latent features and $\mathbf{y}$ are the subject demographics. During training, we decorrelate the latent state $\mathbf{z}=E_\phi(\mathbf{x})$ from demographic features $\mathbf{y}$ so that all of the fMRI signal that can be attributed to demographics is based on user-provided input and not on the encoded latent features. To this end, we make several modifications to the traditional VAE loss function.

\subsubsection{Incorporate Demographic Information}

First, the reconstruction error term of the loss function remains unchanged from the ELBO formulation, except for the injection of demographic information:
\begin{equation}
    \mathcal{L}_{\text{Recon}} = \frac{1}{N}\sum^N_{i=1}\lVert \mathbf{x}_i-D_\theta(\mathbf{z}_i,\mathbf{y}_i)\rVert^2_2,
\end{equation}
where $N$ is the number of subjects, $\mathbf{x}_i$ are the vectorized FC features, $\mathbf{z}_i=E_\phi(\mathbf{x}_i)$ are the empirically calculated latent features, and $\mathbf{y}_i$ are the subject demographics for subject $i$. 

\subsubsection{Extension to Multidimensional Latent Space}

Second, we note that the ELBO loss function of the standard VAE is applicable to scalar latent features $z$ and not multi-dimensional latent features $\mathbf{z}\in\mathbb{R}^{N_z}$. This may allow for a non-diagonal covariance matrix in the empirical distribution of latents $q_\phi(\mathbf{z}|\mathbf{x})=\mathcal{N}(\mathbf{0},\mathbf{\Sigma})$. Thus, we modify a part of the ELBO loss function to specifically target a diagonal covariance matrix and zero expected value for the latents:
\begin{equation}
    \begin{split}
        \mathcal{L}_{\text{Cov}} &= \frac{1}{N}\lVert \mathbf{Z}\mathbf{Z}^\top - N\mathbf{I} \rVert^2_\text{F} \\
        \mathcal{L}_{\text{Mean}} &= \frac{1}{NN_z}\sum^{N_z}_{i=1}\lVert \mathbf{\mu}_{\mathbf{z}_i} \rVert^2_2, 
    \end{split}
\end{equation}
where $\mathbf{Z}\in\mathbb{R}^{N_z\times N}$ is the matrix of all $N_z$ latent features for all $N$ subjects, $\mathbf{z}_i$ is the vector of latent feature $i$ for all $N$ subjects, and $\mathbf{\mu}_{\mathbf{z}_i}$ is its mean. We find that this loss function performs as good or better than the KL divergence part of ELBO with fMRI data.

\subsubsection{Decorrelate Latent Features from Demographics}

Third, we add a term penalizing correlations between the empirical latent features and four demographic features or clinical outcomes: age, sex, race, and disease status (schizophrenia diagnosis). Where we have multiple fMRI scans using different scanner tasks for each subject, we also decorrelate the latents with respect to scanner task. We define
\begin{equation}
    \label{eq:demo}
    \mathcal{L}_{\text{Demo}}= \frac{1}{N_zN_y}\sum^{N_z}_{j=1}\sum^{N_y}_{k=1} \lVert \rho_{\mathbf{z}_j,\mathbf{y}_k} \rVert^2_2,
\end{equation}
where $\rho_{\mathbf{z}_j,\mathbf{y}_k}$ is the correlation between between latent feature $\mathbf{z}_j$ and demographic feature $\mathbf{y}_k$ across all $N$ subjects. 

\subsubsection{Classifier Guidance}
Finally, while training the DemoVAE, we create synthetic samples based on random choices of demographic inputs, and penalize miss-predictions relative to pre-trained models. Given a single demographic prediction from a synthetic latent based on user-input demographics $\hat{y}_i=f_i(D_\theta(\mathbf{z}_\text{samp},\mathbf{y}))$, we define
\begin{equation}
    \label{eq:pred}
    \begin{split} 
        \mathcal{L}_{\text{Guide}} &= \frac{1}{N_y}\sum^{N_y}_{i=1} \begin{cases}
             \lVert y_i - \hat{y}_i \rVert^2_2, 
             & \mathbf{y}_i \;\; \text{continuous}
             \\
              - \sum_{c}y_{i,c}\text{log}(p_{i,c}),
             & \mathbf{y}_i \;\; \text{categorical}
        \end{cases}
    \end{split}
\end{equation}
where the models $f_i(\cdot)$ are linear models trained on the ground truth fMRI subject data, $y_{i,c}$ is the one hot encoded true class label for demographic $i$, $p_{i,c}$ is the predicted probability for class $c$ and demographic $i$, and the loss is the Mean Square Error (MSE) for continuous demographics (age) and Cross Entropy (CE) error for categorical demographics (sex, race, disease status, scanner task).

The final loss function for training the DemoVAE can thus be formulated as:
\begin{equation}
    \begin{split}
    \mathcal{L} = \mathcal{L}_{\text{Recon}} + & \lambda_1 \mathcal{L}_{\text{Cov}} + \lambda_2 \mathcal{L}_{\text{Mean}} + \\ & \lambda_3 \mathcal{L}_{\text{Demo}} + \lambda_4 \mathcal{L}_{\text{Guide}},
    \end{split}
\end{equation}
where $\lambda_{1-4}$ are the hyperparameters chosen alongside learning rate and latent dimension size via random grid search.

\subsection{Generation of Timeseries}
\label{subsec:timeseries}

The DemoVAE model described works on fixed length input vectors to produce fixed length latent feature vectors and fixed length samples from the distribution of the input. When creating fMRI-derived samples of synthetic FC, we can use the Cholesky decomposition \cite{Dereniowski2004-sw} to generate variable-length BOLD timeseries that are compatible with the generated FC. These timeseries may then be used to generate alternate measures of connectivity, e.g., partial correlation-based FC (PCFC).

For fixed-length FC input, which is a symmetric positive semi-definite (PSD) matrix, we can train the decoder function $\hat{\mathbf{X}} = D_\theta(\mathbf{z},\mathbf{y})$ to output a form that can be converted into a symmetric matrix. This can be either the unique upper triangular entries of the matrix $\hat{\mathbf{X}}_U$ or a low-rank factor of size $\mathbf{A}=\mathbb{R}^{N_{roi}\times N_{r}}$, where $N_{roi}$ is the number of regions of interest (ROIs) in the atlas and $N_{r}$ is the rank of the factor:
\begin{equation}
    \begin{split}
        \hat{\mathbf{X}}^{(1)} &= \hat{\mathbf{X}}_U + \hat{\mathbf{X}}_U^\top + \mathbf{I} \\
        \hat{\mathbf{X}}^{(2)} &= \mathbf{A}\mathbf{A}^\top.
    \end{split}
\end{equation}

Note that $\hat{\mathbf{X}}^{(1)}$ may contain a few negative eigenvalues while $\hat{\mathbf{X}}^{(2)}$ is most likely rank deficient, based on the choice of $N_r$. However, the Cholesky decomposition requires a positive definite (PD) matrix as input. In the first case, we find a negligible loss of predictive ability by setting negative eigenvalues $\lambda_{\hat{\mathbf{X}},i}$ of $\hat{\mathbf{X}}$ to zero or a small positive value $\beta$. 
\begin{equation}
    \lambda_{\hat{\mathbf{X}},i} = \begin{cases}
        \lambda_{\hat{\mathbf{X}},i}, & \qquad \lambda_{\hat{\mathbf{X}},i} \geq 0 \\
        \beta, & \qquad \text{otherwise}
    \end{cases}
\end{equation}

We then choose the standard deviation of timeseries $\sigma_i\in N(\mu_{\sigma_i},\tau^2_{\sigma_i})$ at each ROI, and use it to recompute the covariance matrix $\mathbf{\Sigma}$ from the reconstructed or synthetic FC sample $\hat{\mathbf{X}}$:
\begin{equation}
    \mathbf{\Sigma} = (\mathbf{1}{\mathbf{\sigma}}^{\top})\hat{\mathbf{X}}(\mathbf{\sigma}\mathbf{1}^{\top}),
\end{equation}
where $\mathbf{\sigma}$ is a column vector of standard deviations at each ROI and $\mathbf{1}$ is a column vector of ones. Given a rank-deficient but PSD covariance matrix, we can simulate a Cholesky-like decomposition using the eigenvectors and eigenvalues of $\mathbf{\Sigma}$ and QR decomposition in the following way:
\begin{equation}
    \begin{split}
        \mathbf{\Sigma} &= \mathbf{V}\mathbf{\Lambda}\mathbf{V}^\top \\
        &= (\mathbf{V}\sqrt{\mathbf{\Lambda}})(\mathbf{V}\sqrt{\mathbf{\Lambda}})^\top \\
        &= (\mathbf{Q}\mathbf{R})^\top
        (\mathbf{Q}\mathbf{R}) \\
        &= \mathbf{L}\mathbf{L}^\top.
    \end{split}
\end{equation}

Timeseries may then be constructed based on the property of the Cholesky decomposition that a standard normal random variable $X\in\mathcal{N}(0,1)$ multiplied by the Cholesky factor $\mathbf{L}$ creates a multivariate normal variable vector with zero mean and covariance matrix $\mathbf{\Sigma}=\mathbf{L}\mathbf{L}^\top$. This is compatible with fMRI data, which are usually bandpass filtered prior to analysis. It assumes, however, that the timeseries BOLD signal is stationary, which is sufficient when producing correlation-based metrics. In Table~\ref{tab:pnc-prediction}, we show that this method of generating PCFC via timeseries yields features that make accurate predictions using models trained on the original data, and vice-versa.

\subsection{Datasets}
\label{subsec:datasets}

We now describe two datasets used for validation and exploration of the DemoVAE model. Demographics for our subsets of the two data sources may be found in Table~\ref{tab:demo}.

\begin{table}
    \centering
    \caption{
    \label{tab:demo}Demographics for the PNC and BSNIP datasets.}
    \begin{tabular}{|l|c|c|c|}
    \multicolumn{4}{c}{PNC Dataset} \\
    \hline
    & Males & Females & p-value \\
    \hline
    Age (years) & $14.48\pm3.32$ & $14.69\pm3.42$ & $<0.3042$ \\
    European Anc. (EA) & 316 & 303 & $<0.5415$ \\
    African Anc. (AA) & 224 & 311 & $<10^{-5}$ \\
    WRAT Score & $103.66\pm16.56$ & $101.45\pm15.90$ & $<0.0212$\\
    \hline
    & EA & AA & \\
    \hline
    Age (years) & $14.59\pm3.50$ & $14.59\pm3.23$ & $<0.9994$ \\
    WRAT Score & $108.68\pm14.84$ & $95.68\pm14.80$ & $<10^{-48}$ \\
    \hline
    \multicolumn{4}{c}{} \\
    \end{tabular}
    \begin{tabular}{|l|c|c|c|}
    \multicolumn{4}{c}{BSNIP Dataset} \\
    \hline
    & Males & Females & p-value \\
    \hline
    Age (years) & $35.24\pm11.98$ & $38.93\pm12.45$ & $<0.0026$ \\
    Caucasian Anc. (CA) & 139 & 109 & $<0.0221$ \\
    African Anc. (AA) & 82 & 75 & $<0.5343$ \\
    SZ Diagnosis & 130 & 55 & $<10^{-10}$ \\
    \hline
    \end{tabular}
\end{table}

\subsubsection{Philadelphia Neurodevelopmental Cohort}
\label{subsubsec:pnc}

The Philadelphia Neurodevelopmental Cohort (PNC) is a widely-used dataset of children and young adults with multi-task fMRI scans for 1,529 subjects \cite{Satterthwaite2014NeuroimagingOT} and genomic data for more than 9,000 \cite{Glessner2010-xg}, many of whom have both modalities. In addition, the PNC includes data for 169 questionnaire, computerized battery, and in-scanner task parameter fields \cite{Calkins2015-lc}\cite{Gur2014}, not all of which are available for every subject. Scholastic achievement was measured using the Wide Range Achievement Test (WRAT) \cite{Sayegh2014QualityOE}, with both a raw score and score with the effects of age regressed out. The dataset is enriched for subjects of European (EA) and African (AA) ancestry. fMRI scans include three in-scanner tasks: a resting state (rest), a working n-back memory (nback) \cite{Ragland2002WorkingMF}, and an emotion identification task (emoid), where not all subjects have all three tasks. We selected a 1,154-subject subset of the entire cohort that included subjects with all three fMRI scanner tasks as well as single nucleotide polymorphism (SNP) data, and who belonged to either of the two predominant ancestry groups. 

Acquisition \cite{Satterthwaite2014NeuroimagingOT} and preprocessing \cite{Orlichenko2023-ue} of the fMRI data has been described previously \cite{10002422}, but was performed using a whole-body 3T scanner running an echo-planar imaging sequence with a repetition time of $\text{TR}=3$sec. Data was pre-processed using SPM12\footnote{\url{http://www.fil.ion.ucl.ac.UK/spm/software/spm12/}}, including regression for motion correction, co-registration, and normalization to MNI space. The Power Atlas\cite{Power2011FunctionalNO} of 5mm spherical regions was used to parcellate the fMRI BOLD images into 264 timeseries. FC was created from these timeseries via Pearson correlation. Partial correlation-based FC was created from these timeseries via the nilearn\footnote{\url{https://nilearn.github.io/stable/index.html}} software package\cite{Abraham2014} using the \textit{Ledoit-Wolf} shrinkage estimator\cite{Ledoit2004-fw}.

SNP data was collected using one of eight different platforms, each subject's data being handled using one of these platforms, with the largest platform annotating 1,185,051 SNPs. For our analysis, we chose a subset of 35,621 SNPs that were available on all 8 platforms for all subjects. SNPs were categorized by haplotypes as homozygous dominant, heterozygous, homozygous recessive, or missing.

\begin{figure}
    \centering
    \hspace{-6mm}\includegraphics[width=0.53\textwidth]{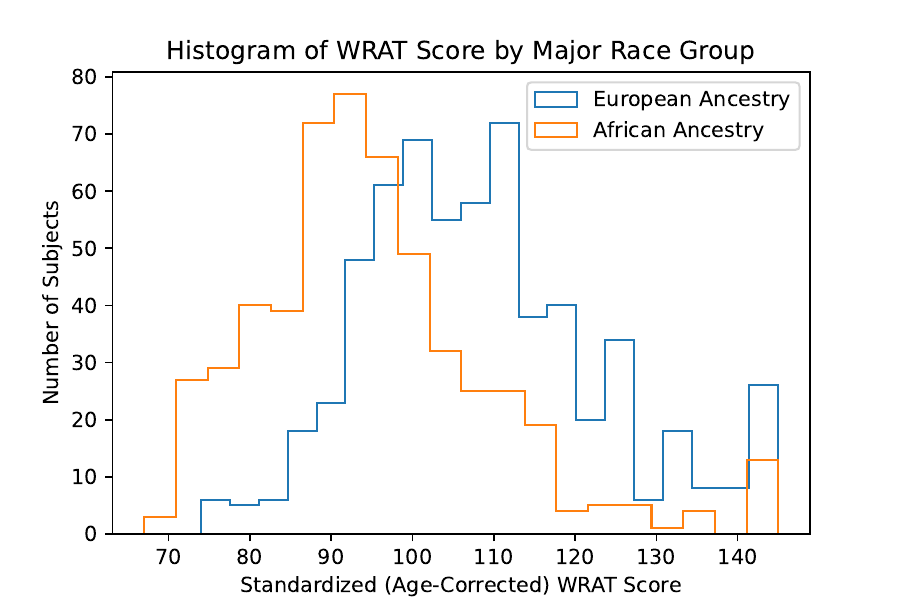}
    \caption{
    \label{fig:wrat-hist}Histogram of standardized (age-correct) WRAT score from the PNC dataset, split among the two major race groups in the dataset. There is a clear demographic confound when predicting WRAT score from fMRI or genomic data. We show in Table~\ref{tab:wrat-pred} that DemoVAE is able to remove the effect of this confound, but at the same time, removes the ability to accurately predict WRAT score.}
\end{figure}

We note the WRAT scores with age regressed out in Table~\ref{tab:demo} have not been adjusted for race, as seen from the very significant p-value, implying the possibility of confounding effects. Figure~\ref{fig:wrat-hist} displays the histogram of WRAT score among the two races. One of the goals of the DemoVAE model is to remove the effect of demographics from fMRI features in order to give an unconfounded view of the effect of brain network organization on phenotypic variables, e.g., removing the effects of demographics on scholastic achievement score. 

\subsubsection{Bipolar and Schizophrenia Network for Intermediate Phenotypes}
\label{subsubsec:bsnip}

As an additional and independent validation dataset containing clinical phenotypes, we use the Bipolar and Schizophrenia Network for Intermediate Phenotypes cohort of 933 patients, 1059 relatives, and 459 healthy controls \cite{Tamminga2014-pp}. We selected a subset of 185 schizophrenia (SZ) patients and 220 healthy controls for whom we had fMRI scans, excluding subjects with borderline diagnosis such as bipolar and schizoaffective disorder. fMRI data was acquired over six sites, with acquisition and preprocessing of the data described elsewhere \cite{8857902}. In addition to the fMRI data, the BSNIP dataset contains 32 medication and clinical assessment measures related to patients' psychiatric condition.

\subsection{Experiments}
\label{subsec:experiments}

\subsubsection{Prediction of WRAT Score Using DemoVAE Latents}

We used the PNC dataset to predict age-adjusted WRAT score, which is heavily skewed according to ethnic group (Figure~\ref{fig:wrat-hist}), using fMRI FC data, SNP data, scalar race indicator, and DemoVAE latents constructed from FC or SNPs. Ridge regression models were trained and evaluated on a set of 20 repetitions of an 80/20 train/test split with the above features, where the best value for the regularization parameter was chosen by random grid search. This experiment was performed to validate the ability of DemoVAE to decorrelate its latent features from demographics, and to demonstrate why demographic confounds in FC may be problematic for downstream analysis.

\subsubsection{Validation of fMRI Samples Generated by DemoVAE}
\label{subsubsec:synthetic_validation}

Several tests were performed to validate that the samples created by DemoVAE accurately capture the distribution of fMRI data and recapitulate group differences between groups having different demographics. We first trained the DemoVAE model using the PNC dataset, including age, sex, and race as demographics, and with the scanner task being set to resting state. We also trained a traditional VAE using the traditional scalar ELBO objective in Equation~\ref{eq:elbo_loss} and no demographic information, as well as a Wasserstein generative adversarial network (W-GAN) model \cite{goodfellow2014generative}\cite{pmlr-v70-arjovsky17a}\cite{Laino2022-gy}. Synthetic FC samples were then generated for 1,000 subjects using all three models, and the distribution of FC features was visualized in two dimensions using the scikit-learn implementation of t-distributed stochastic neighbor embedding (t-SNE) \cite{pedregosa2011scikit}\cite{JMLR:v9:vandermaaten08a}. Subject demographics for the DemoVAE features were sampled randomly using an equally-weighted Bernoulli (sex, race) or normal (age) distribution.  The distribution of synthetic data was compared with ground truth data.

Additionally, we measured the ability of DemoVAE synthetic data to recapitulate group differences in the PNC and BSNIP datasets. We calculated the mean difference in FC between young children and young adults, males and females, EA and AA race, and SZ patients and healthy controls using ground truth data. Then, we created synthetic FC data for those groups using using DemoVAE, and compared group differences of real and synthetic data. The RMSE between FC differences of real and synthetic data was calculated and compared with a null model. 

\subsubsection{Phenotype Prediction Using DemoVAE Synthetic Data}

The ability of DemoVAE to create synthetic data that recapitulate the demographic content of subject FC was tested by using real data to train demographic-prediction models that were tested on synthetic data and vice versa. The models used were Ridge regression models for continuous variables (age) and Logistic regression models for binary variables (sex, race, SZ diagnosis). The scikit-learn implementation of these models were employed and optimal regularization parameters were chosen using random grid search. Synthetic data was created using the same procedure as in Section~\ref{subsubsec:synthetic_validation}. Twenty repetition of each experiment was performed and the results averaged. The same number of synthetic subjects were created as available real subjects: 1,154 for the PNC dataset and 405 for BSNIP.

\subsubsection{Correlation of Clinical Measures with DemoVAE Latents}

We tested the correlation of fMRI FC data with phenotype and clinical data fields before and after the removal of the confounding effects of demographics. Both the PNC and BSNIP dataset contain phenotype and clinical data which may be correlated with FC features. A subset of 169 phenotype, medication, and cognitive battery fields available in the PNC cohort was correlated with raw FC data, traditional VAE latents, and DemoVAE latents decorrelated from demographic features. Correlation was tested at a significance level of $p<0.05$ and $p<0.01$, and the number of significant correlations was determined. Significance was determined using a t-test with the statistic:
\begin{equation}
    t=\frac{\rho\sqrt{n-2}}{\sqrt{1-\rho^2}},
\end{equation}
where $\rho$ was the correlation coefficient between FC or latent feature and clinical or computerized battery field and $n$ was the number of samples, i.e., number of subjects having a value for that clinical or computerized battery field. Each FC, VAE, or DemoVAE feature was correlated independently and Bonferroni correction was applied to the p-value to correct for multiple comparisons.

In additional to the PNC clinical fields, the BSNIP dataset contained 32 demographic, clinical, and medication fields which were correlated with FC data and VAE latent features in a similar manner. Finally, the PNC dataset contained genomic data for a 1,154-subject subset of subjects with fMRI scans. These genomic data were also correlated with phenotype and cognitive battery fields before and after removal of confounding effects with DemoVAE. 

\subsubsection{Imputation of fMRI Scanner Task}

DemoVAE creates latent features that are decorrelated from fMRI scanner task, and can generate samples conditioned on the type of scanner task. We therefore test the ability of DemoVAE to impute scanner task fMRI given fMRI from a different scanner task as input. Imputation was performed either deterministically, by switching the identity of the task $y_i$ the decoder $D_\theta(\mathbf{z},\mathbf{y})$ was conditioned on, or by switching task in addition to adding 10\% noise in the latent dimension $\mathbf{z}$. Imputation accuracy using DemoVAE, as measured by RMSE, was compared to using uniformly zero FC, the input scanner task, adjusting by the mean difference between tasks in the training set, and a two-layer MLP model.

\section{Results}
\label{sec:results}

This section presents results for the experiments described in Section~\ref{subsec:experiments}.

\subsection{Prediction of WRAT Score from Decorrelated Latents}
\label{subsec:wrat}

In Table~\ref{tab:wrat-pred}, we give results for predicting age-adjusted WRAT score in the PNC data from scalar race value, FC data, SNP data, DemoVAE latents derived from FC data, and DemoVAE latents derived from SNPs. We observe that using the scalar race variable yields the best prediction of standardized WRAT score. While FC and SNPs can predict WRAT score moderately well, that predictive ability disappears when latents are decorrelated from race, as in the DemoVAE latents. This demonstrates that DemoVAE is able to decorrelate the fMRI latent state from demographics. It also demonstrates that, while FC and SNPs have the ability to predict age-adjusted WRAT score, that prediction is based on ability to infer demographics, and not on any cognitive signal found in FC that is independent of demographics. We find, as have previous studies, that prediction of scholastic achievement may be highly confounded by race signal present in neuroimaging data \cite{Li2022-xf}\cite{Orlichenko2023-ue}.

\begin{table}
    \centering
    \caption{
    \label{tab:wrat-pred}RMSEs (mean and standard deviation) of predicting standardized WRAT scores using fMRI FC input, SNP input, DemoVAE fMRI latents, DemoVAE SNP latents, and scalar race variable.}
    \begin{tabular}{|l|c|}
        \hline
         Input & WRAT Prediction RMSE \\
         \hline
         \hline
         Null Model & $15.18$ \\
         \hline
         Race Only & $\mathbf{13.91\pm0.271}$ \\
         \hline
         Rest FC & $14.73\pm0.368$ \\
         Nback FC & $14.44\pm0.395$ \\
         Emoid FC & $14.46\pm0.414$ \\
         SNPs & $14.03\pm0.429$ \\
         \hline
         Rest DemoVAE Latents & $15.20\pm0.015$ \\
         Nback DemoVAE Latents & $15.18\pm0.013$ \\
         Emoid DemoVAE Latents & $15.18\pm0.015$ \\
         SNP DemoVAE Latents & $15.14\pm0.131$ \\
         \hline
    \end{tabular}
\end{table}

\subsection{Validation of fMRI Samples Generated by DemoVAE}

\begin{figure}
    \centering
    \includegraphics[width=0.49\textwidth]{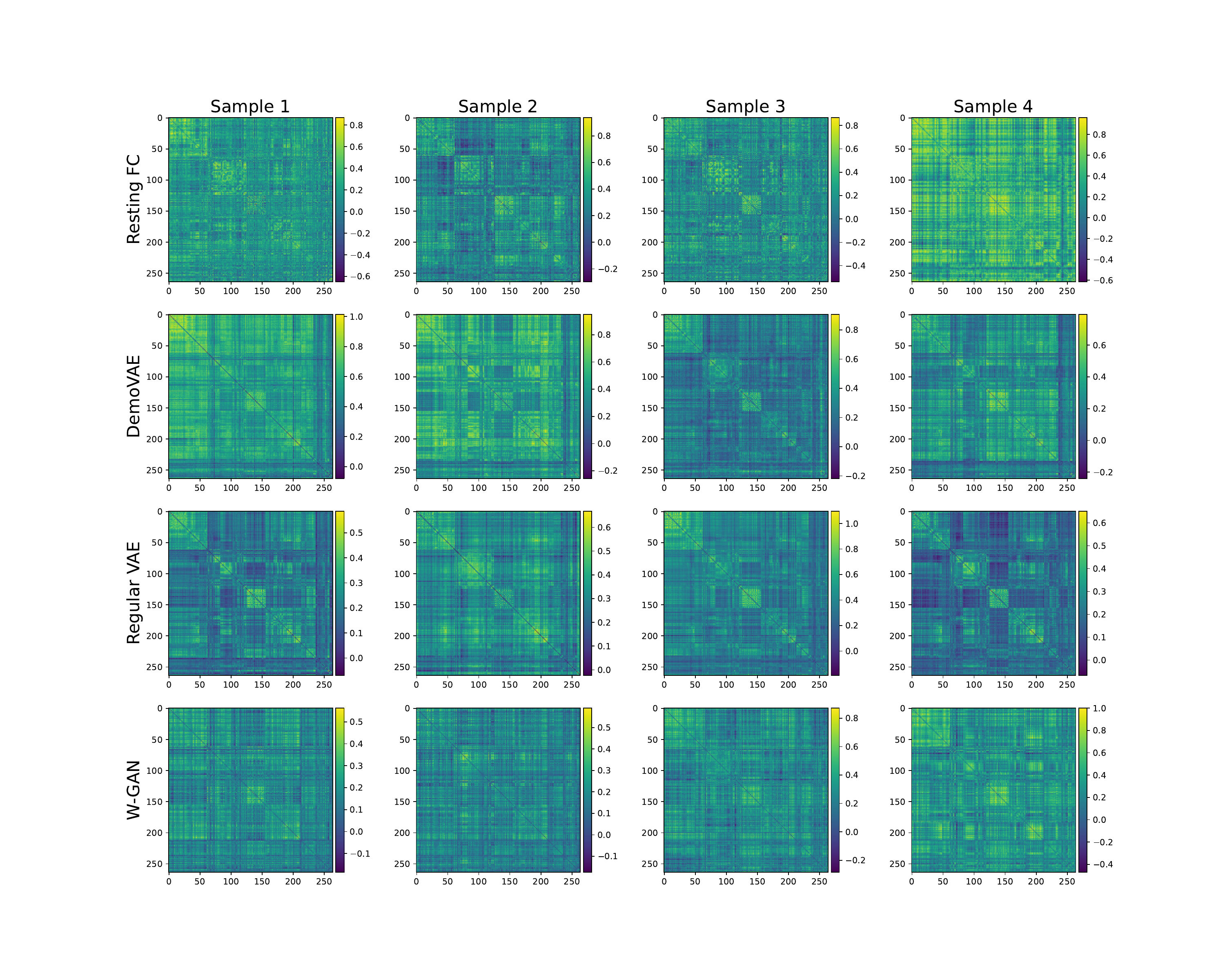}
    \caption{
    \label{fig:samples}Sampled FC matrices for real PNC resting state scans (top) compared to synthetic DemoVAE, VAE, and W-GAN FC data. Visually, all synthetic models generate convincing data.}
\end{figure}

\begin{figure*}
    \centering
    \includegraphics[width=\textwidth]{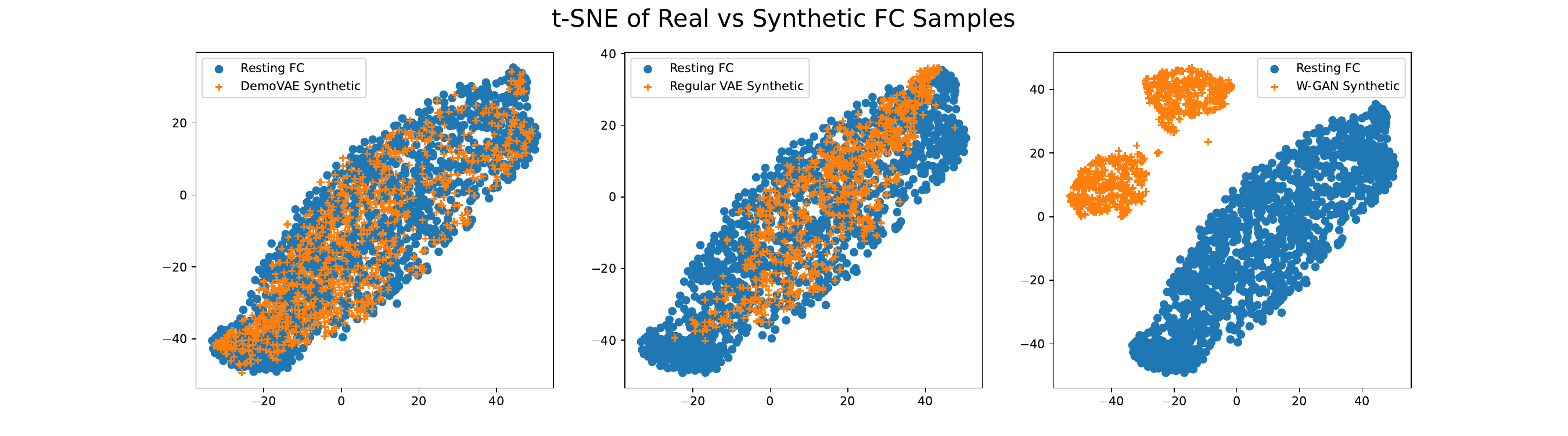}
    \caption{
    \label{fig:tsne}t-SNE embeddings of synthetic FC data from DemoVAE, traditional VAE, and W-GAN models overlayed on top of t-SNE embeddings of real resting state FC data from the PNC dataset. Blue circles represent embeddings of real subject FC data while orange crosses represent embeddings of synthetic data. We see that DemoVAE captures the distribution of fMRI FC data as well as or better than a traditional VAE and better than a GAN.}
\end{figure*}

Figure~\ref{fig:samples} displays a selection of ground truth subject FC data compared to synthetic data generated by DemoVAE, a traditional VAE, and a W-GAN. We note that it is visually hard to distinguish between true subject data and synthetic data. However, this is not the case when comparing the entire distribution of data using t-SNE, as evident in Figure~\ref{fig:tsne}. Figure~\ref{fig:tsne} shows the distribution of synthetic DemoVAE data, VAE data, and W-GAN data transformed using t-SNE overlayed on ground truth resting state PNC subject data. DemoVAE data was created using randomly sampled age, sex, and race demographics but with scanner task set to resting state. We see that DemoVAE captures the distribution of fMRI data better than the traditional VAE and W-GAN. It is evident that a GAN makes no guarantees about matching or even approximating the true distribution of data \cite{pml2Book} unless additional regularization is performed.

\begin{figure*}
    \centering
    \includegraphics[width=18cm]{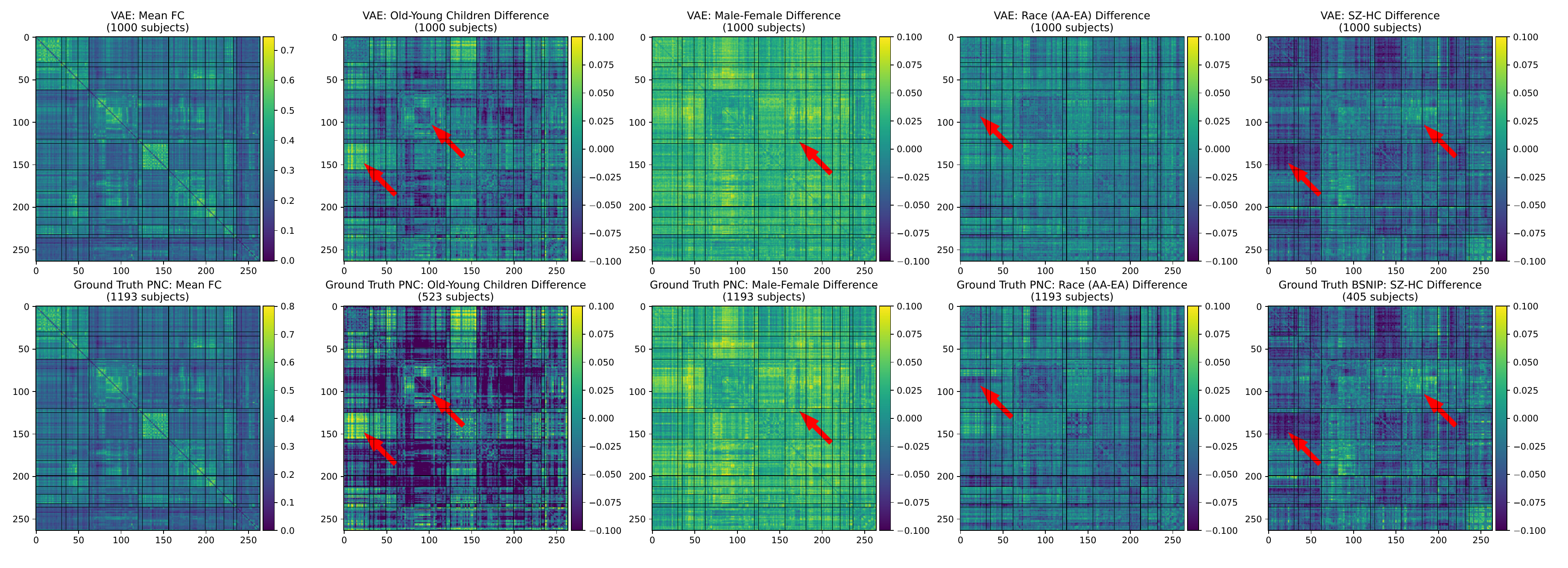} 
    \caption{
    \label{fig:groupdiff}Group FC differences using real data and synthetic data generated by DemoVAE conditioned on appropriate demographic input. Top: synthetic DemoVAE data, bottom: real data. From left to right, we see that DemoVAE qualitatively recapitulates group differences in the PNC (mean, age, sex, race) and BSNIP (SZ diagnosis) datasets. Arrows point out FC features in real data that are reproduced in synthetic DemoVAE samples. Brain functional networks for the Power atlas, shown left to right and top to bottom in FC matrices, are given in Table~\ref{tab:bfns}.}
\end{figure*}

Figure~\ref{fig:groupdiff} displays group differences between demographic subsets of real data compared to group differences from synthetic DemoVAE data. We see that by conditioning on demographic input, DemoVAE can produce samples that accurately recapitulate group differences in FC data. Table~\ref{tab:groupdiff} shows RMSE values for deviation in group differences between synthetic DemoVAE data and real data.

\begin{table}
    \centering
    \caption{\label{tab:bfns}Brain functional networks in the Power264 atlas \cite{Power2011FunctionalNO}.}
    \begin{tabular}{|c|c|l|}
        \hline
        Label & ROIs & Network Name \\
        \hline
        \hline
         0 & 0-29 & Somatomotor Hand (SMT) \\
         \hline
         1 & 30-34 & Somatomotor Mouth (SMT) \\  \hline
         2 & 35-48 & Cinguloopercular (CNG) \\  \hline
         3 & 49-61 & Auditory (AUD) \\  
         \hline
         4 & 62-119 & Default Mode (DMN) \\  \hline
         5 & 120-124 & Memory (MEM) \\ \hline
         6 & 125-155 & Visual (VIS) \\ \hline
         7 & 156-180 & Frontoparietal (FRNT) \\ \hline
         8 & 181-198 & Salience (SAL) \\ \hline
         9 & 199-211 & Subcortical (SUB) \\ \hline
         10 & 212-220 & Ventral Attention (VTRL) \\ \hline
         11 & 221-231 & Dorsal Attention (DRSL) \\ \hline
         12 & 232-235 & Cerebellar (CB) \\ \hline
         13 & 236-263 & Uncertain (UNK) \\ \hline
    \end{tabular}
\end{table}

\begin{table}
    \centering
    \caption{
    \label{tab:groupdiff}RMSEs between FC group differences using real versus synthetic DemoVAE data.}
    \begin{tabular}{|l|c|c|}
        \hline
         Groups & RMSE & RMSE \\
         & Real/Synthetic & Real/Null Model \\
         \hline
         \hline
         Mean of All & $\mathbf{0.0353}$ & $0.3091$ \\
         \hline
         Young/Old Children & $\mathbf{0.0336}$ & $0.0717$ \\
         \hline
         Males/Females & $\mathbf{0.0117}$ & $0.0379$ \\
         \hline
         EA/AA Race & $\mathbf{0.0194}$ & $0.0359$ \\
         \hline
         SZ/Healthy Controls & $\mathbf{0.0090}$ & $0.0453$ \\
         \hline
    \end{tabular}
\end{table}

\subsection{Phenotype Prediction Using DemoVAE Synthetic Data}

Table~\ref{tab:pnc-prediction} shows the predictive RMSE and accuracy when training models on real fMRI data and predicting on synthetic DemoVAE data and vice versa. Predictive tasks include age, sex, race, and SZ diagnosis prediction. The predictive accuracy is very high when training using real data and predicting using DemoVAE, and slightly lower when training using DemoVAE and predicting using real data, but still exceeds 90\% in all but one instance. Pearson FC and partial correlation-based FC derived using the FC-based timeseries creation procedure described in Section~\ref{subsec:timeseries} have similar accuracies. This validates our timeseries creation procedure, at least in the context of calculation of alternate measures of connectivity.

\begin{table*}
\centering
    \caption{
    \label{tab:pnc-prediction}Transfer of models between fMRI and VAE. RMSE (age prediction) and mean accuracy (sex, race, and schizophrenia prediction) for MLP models trained on ground truth fMRI data and tested on DemoVAE generated samples and vice versa. FC=Pearson Functional Connectivity, PCFC=Partial Correlation-based Functional Connectivity}
    \begin{tabular}{|l|c|c|c|c|c|c|c|}
        \multicolumn{8}{c}{Train on fMRI, Test on VAE} \\
        \multicolumn{8}{c}{PNC Dataset} \\
        \hline
        Predictive Task & Null Model & Rest FC & Rest PCFC & Nback FC & Nback PCFC & Emoid FC & Emoid PCFC \\  
        \hline
        Age (years, RMSE) & 3.30 & \textbf{0.570} & 2.181 & \textbf{0.468} & 1.97 & \textbf{0.495} & 1.91 \\
        \hline
        Sex (ACC, \%) & 53.2 & \textbf{100} & \textbf{99.7} & \textbf{100} & \textbf{99.6} & \textbf{99.9} & \textbf{99.4} \\
        \hline
        Race (ACC, \%) & 53.0 & \textbf{100} & \textbf{99.8} & \textbf{100} & \textbf{99.9} & \textbf{100} & \textbf{100} \\
        \hline
        \multicolumn{8}{c}{}
    \end{tabular}

    \begin{tabular}{|l|c|c|c|c|c|c|c|}
        \multicolumn{8}{c}{Train on VAE, Test on fMRI} \\
        \multicolumn{8}{c}{PNC Dataset} \\
        \hline
        Predictive Task & Null Model & Rest FC & Rest PCFC & Nback FC & Nback PCFC & Emoid FC & Emoid PCFC \\  
        \hline
        Age (years, RMSE) & 3.30 & 2.032 & 2.752 & 1.848 & 2.567 & 1.953 & 2.597 \\
        \hline
        Sex (ACC, \%) & 53.2 & 88.9 & 90.4 & 90.9 & 91.2 & 91.1 & 91.7 \\
        \hline
        Race (ACC, \%) & 53.0 & 93.2 & \textbf{96.3} & 93 & \textbf{96.1} & 93.1 & \textbf{97.1} \\
        \hline
        \multicolumn{8}{c}{}
    \end{tabular}
    
    \begin{tabular}{|l|c|c|c|}
        \multicolumn{4}{c}{Train on fMRI, Test on VAE} \\
        \multicolumn{4}{c}{BSNIP Dataset} \\
        \hline
        Predictive Task & Null Model & FC & PCFC \\  
        \hline
        Age (years, RMSE) & 12.4 & \textbf{3.67} & \textbf{3.86} \\
        \hline
        Sex (ACC, \%) & 54.5 & \textbf{100} & \textbf{100} \\
        \hline
        Race (ACC, \%) & 61.2 & \textbf{100} & \textbf{100} \\
        \hline
        Schizophrenia (ACC, \%) & 54.3 & \textbf{100} & \textbf{98.7} \\
        \hline
        \multicolumn{4}{c}{}
    \end{tabular}
    \begin{tabular}{|l|c|c|c|}
        \multicolumn{4}{c}{Train on VAE, Test on fMRI} \\
        \multicolumn{4}{c}{BSNIP Dataset} \\
        \hline
        Predictive Task & Null Model & FC & PCFC \\  
        \hline
        Age (years, RMSE) & 12.4 & 7.98 & 10.3 \\
        \hline
        Sex (ACC, \%) & 54.5 & \textbf{97.5} & 94.5 \\
        \hline
        Race (ACC, \%) & 61.2 & \textbf{96.0} & 93.3 \\
        \hline
        Schizophrenia (ACC, \%) & 54.3 & 93.3 & 92.3 \\
        \hline
        \multicolumn{4}{c}{}
    \end{tabular}
\end{table*}

\subsection{Correlation of Clinical Measures with DemoVAE Latents}

\begin{figure*}
    \centering
    \includegraphics[width=18cm]{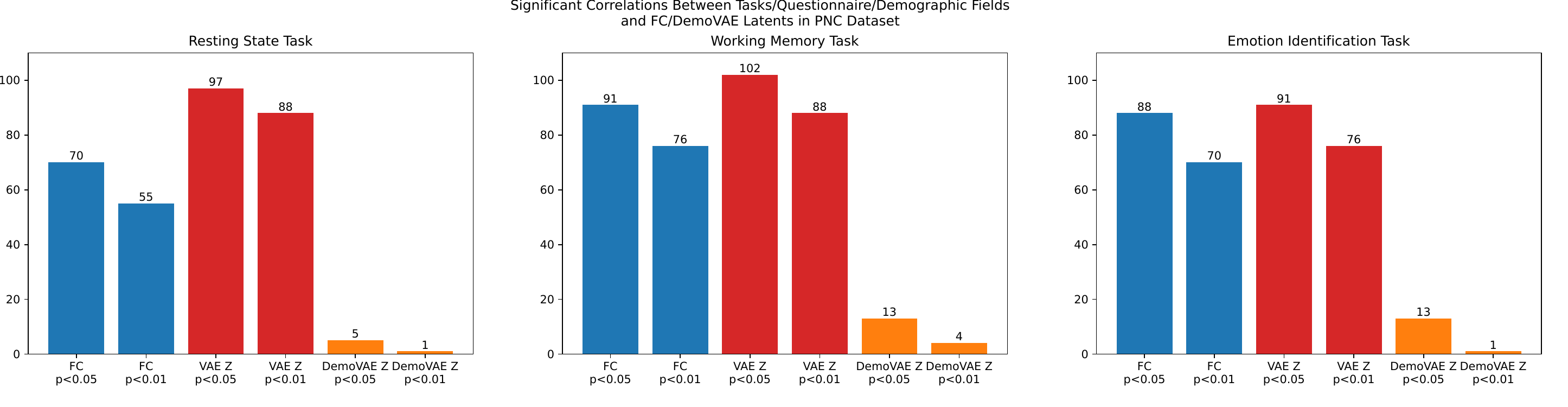}
    \includegraphics[width=6.5cm]{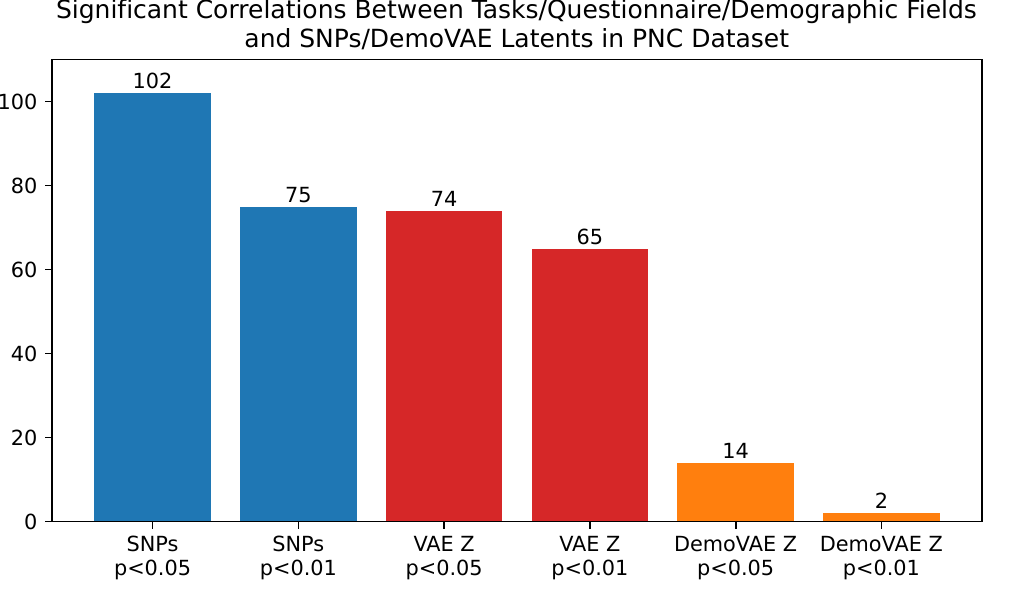}
    \includegraphics[width=6.5cm]{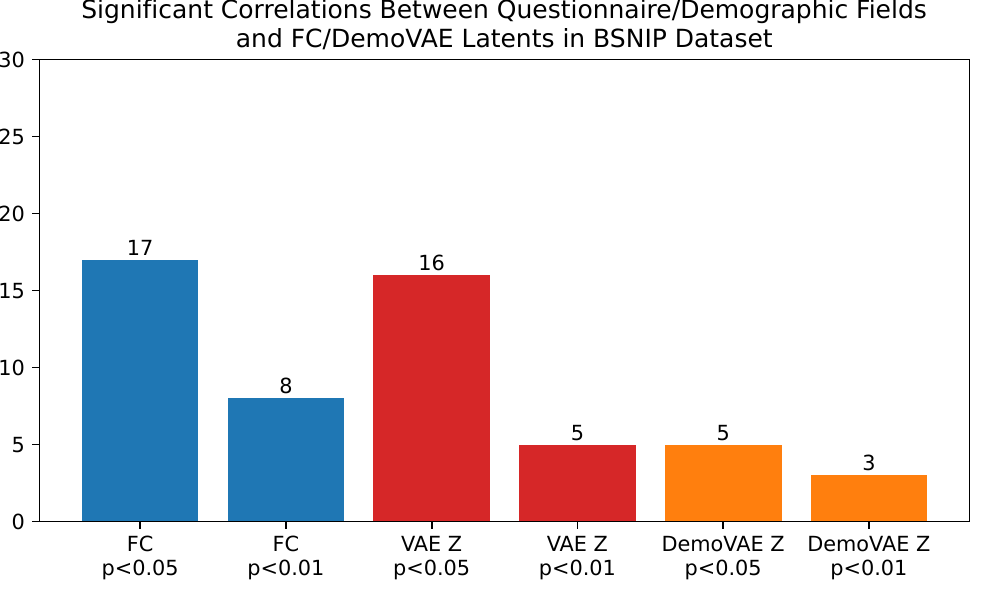}
    \caption{Correlation of questionnaire, computerized battery, and clinical fields with fMRI FC data versus traditional VAE or decorrelated DemoVAE latent features. Top: PNC dataset fMRI data, left bottom: PNC dataset SNP data, right bottom: BSNIP dataset fMRI data. There were a total of 169 fields in the PNC dataset and 32 in the BSNIP dataset. We see that both FC and traditional VAE latents, which are confounded by patient demographics, have significant correlations with more than half of all fields. Once demographic confounds are removed with DemoVAE, however, both FC and SNP data are significantly correlated with only a small percentage of fields. A list of the fields used is available in the GitHub repository. Blue color=correlation with FC features, Red color=correlation with regular VAE latents, Orange color=correlation with DemoVAE latents}
    \label{fig:zcorr}
\end{figure*}

Figure~\ref{fig:zcorr} displays the correlation between clinical questionnaire and computerized battery fields of the PNC and BSNIP datasets and fMRI FC data, traditional VAE latents, and demographically-unconfounded DemoVAE latents. We see that removing the effects of demographic confounds from either fMRI data or SNP data greatly reduces the number of fields that are significantly correlated with the fMRI or genomic data. In fact, of 169 clinical or computerized battery fields, only four remained significantly correlated at the $p<0.01$ level after decorrelation from demographics. This result corroborates the result presented in Section~\ref{subsec:wrat}, where it was found that scalar race value was the best predictor of scholastic achievement as measured by WRAT score. While FC and SNPs were found to be somewhat predictive of WRAT score, that predictive ability disappeared when FC features were decorrelated from the demographics age, sex, and race using DemoVAE.

Unlike the PNC dataset, from which we used 169 questionnaire and computerized battery fields, the BSNIP dataset included a more modest 32 clinical fields available for analysis. All fields including descriptions are available at the GitHub repository accompanying this manuscript. When processing BSNIP data with DemoVAE, we used age, sex, race, and schizophrenia diagnosis as demographic variables to decorrelate latent features. Interestingly, the five BSNIP fields that remained correlated to DemoVAE latent features at a significance of $p<0.05$ were related to medication (taking or not taking anti-psychotics, $p<0.0218$) or Positive and Negative Syndrome Scale (PANSS) assessment as to the severity of schizophrenia symptoms \cite{Liechti2017-fv}. These included total positive symptom score ($p<0.0098$), total negative symptom score ($p<0.0296$), total general symptom score ($p<0.0011$), and total PANSS score ($p<0.00033$). This seems to imply that type or severity of schizophrenia symptoms \cite{Dabiri2022-eg} may have effects in fMRI data which are not accounted for by a simple binary diagnosis of the condition or demographics.

\subsection{Imputation of fMRI Scanner Task}
\label{subsec:imputation}

Table~\ref{tab:task-transfer} displays RMSEs when imputing FC from one fMRI scanner task, i.e., resting state, working memory, or emotion identification, to another. We find incorporating the average training set difference only improved RMSE marginally from simply reusing the input. Either an MLP or DemoVAE in deterministic mode, where only the scanner task ``demographic'' was changed in the decoder $D_\theta(\mathbf{z},\mathbf{y})$, gave approximately the same RMSE, which was significantly better than adding the average of training set difference. By introducing 10\% noise to the latent features created by the DemoVAE encoder $E_\phi(\mathbf{x})$, the RMSE was significantly reduced compared to MLP or deterministic DemoVAE when taking the best of 10 samples. Interestingly, in this case the average error of the 10 samples was not increased significantly compared to MLP or DemoVAE in deterministic mode. These results suggest that there is a wide range of natural variability in FC, even when considering the same subject \cite{Mantwill2022-zw}.

\begin{table*}
\centering
    \caption{
    \label{tab:task-transfer}RMSEs (mean and standard deviation) for the reconstruction of one task FC from another scanner task in the test set, using MLP model, mean difference on training set, and DemoVAE. DemoVAE is used in deterministic mode and using best and average of 10 samples adding 10\% noise in the latent dimension. The ability of the DemoVAE to sample the distribution in the latent space allows it to generate more accurate samples when the transfer function is non-deterministic.}
    \begin{tabular}{|l|c|c|c|c|c|c|}
    \hline
    & Rest $\rightarrow$ Nback & Rest $\rightarrow$ Emoid & Nback $\rightarrow$ Rest & Nback $\rightarrow$ Emoid & Emoid $\rightarrow$ Rest & Emoid $\rightarrow$ Nback \\
    \hline
    Zero FC & $0.333\pm0.091$ & $0.344\pm0.100$ & $0.368\pm0.103$	& $0.344\pm0.100$ & $0.368\pm0.103$ & $0.333\pm0.091$ \\
    \hline
    Reuse Input & $0.218\pm0.058$ & $0.234\pm0.064$ & $0.232\pm	0.059$ & $0.197\pm0.054$ & $0.234\pm0.064$ & $0.194\pm0.055$ \\
    \hline
    Average Training Set Diff. & $0.210\pm0.057$ & $0.231\pm0.064$	& $0.230\pm0.059$ & $0.194\pm0.055$ & $0.231\pm0.064$ & $0.197\pm0.054$ \\
    \hline
    MLP & $0.191\pm0.047$ & $0.200\pm0.050$ & $0.216\pm0.047$ & $\mathbf{0.190\pm0.048}$	& $0.217\pm0.050$ & $\mathbf{0.180\pm0.039}$ \\
    \hline
    DemoVAE (Deterministic) & $0.196\pm0.0526$ & $0.202\pm0.055$	& $0.220\pm0.056$ & $\mathbf{0.188\pm0.048}$ & $0.219\pm0.059$	& $\mathbf{0.183\pm0.047}$ \\
    \hline
    DemoVAE (Best of 10) & $\mathbf{0.176\pm0.028}$ & $\mathbf{0.185\pm0.034}$ & $\mathbf{0.203\pm0.034}$ & $\mathbf{0.185\pm0.032}$ & $\mathbf{0.203\pm0.034}$ & $\mathbf{0.178\pm0.029}$ \\
    \hline
    DemoVAE (Avg. of 10) & $0.197\pm0.052$ & $0.202\pm0.056$ & $0.220\pm0.056$ & $\mathbf{0.189\pm0.048}$ &	$0.219\pm0.059$ & $\mathbf{0.183\pm0.047}$ \\
    \hline
    \end{tabular}
\end{table*}

\section{Discussion}
\label{sec:discussion}

Previous work has found group differences in FC between children and young adults as well as other demographic groups. \textit{Sanders et al.} have identified that somatomotor-visual network resting state functional connectivity in the Human Connectome Project dataset \cite{Van_Essen2012-mj} is most highly correlated with age of child or adolescent \cite{Sanders2023-xc}. Other investigators have found that somatomotor-visual network connectivity showed an increase in connectivity strength in a longitudinal subset of older adults \cite{Orlichenko2023-zv} from the UK Biobank \cite{Sudlow2015-dz}. These data support our finding, shown in Figure~\ref{fig:groupdiff} (second FC matrices from left, leftmost red arrow), of large somatomotor-visual network connectivity differences between older and younger children. \textit{Ficek-Tani et al.} have found sex-related differences in the default mode network (DMN), with females having higher intra-DMN connectivity and males having higher connectivity between DMN and other regions \cite{Ficek-Tani2023-tp}\cite{Jung2015-bc}. This finding is again reproduced by our own simple analysis of PNC data shown in Figure~\ref{fig:groupdiff} (middle FC matrix). While the effects of ethnicity on fMRI have been less widely studied, it has been reported that race may have a large effect on the features of FC data \cite{Li2022-xf}\cite{Orlichenko2023-ue}. Concerning schizophrenia, \textit{Li et al.} have reported significant hypoconnectivities in multiple brain networks \cite{Li2019-jw}, including the somatomotor network, which aligns with our differential FC map in Figure~\ref{fig:groupdiff} (far right FC matrix, left arrow). \textit{Bernard et al.} specifically identified motor networks as contributing to schizophrenia endophenotype \cite{Bernard2017-ks}. The fact that our DemoVAE model is able to reproduce these group differences in synthetic data while capturing the wide variation in individual fMRI data (see Figure~\ref{fig:tsne}) makes it suitable for exploratory use by researchers who do not have permission or have not yet applied to access clinical fMRI datasets. This is further supported by the results, shown in Table~\ref{tab:pnc-prediction}, that models trained on synthetic DemoVAE data perform comparably to models trained on real fMRI data.

Moreover, previous researchers \cite{Li2022-xf}\cite{Orlichenko2023-ue} have highlighted the possibility of prediction based on fMRI data being confounded by demographics, e.g. ethnicity in the prediction of scholastic achievement. Likewise, it is known that the prevalence of schizophrenia may be elevated in men compared to women \cite{Li2022-dw}, or at least that the age of onset of the disease tends to be different in men versus women \cite{Li2016-yx}. In fact, if not regressing out the effects of age, most measures of scholastic achievement would be highly confounded by children's grade level. We believe the ability to generate fMRI latent features where the confounding effects of demographics are removed may be a valuable addition to the analysis of fMRI FC data. Alternately, given the present finding of high confounding effects of demographics in FC data, it may be useful for researchers to begin to consider other and newer modalities, such as FNIRS \cite{Pinti2020-wa}, MEG \cite{Singh2014-gf}, or electrode recordings \cite{Saha2021-iw} in addition to fMRI. 

Although we find significant reductions in correlations with clinical questionnaire or computerized battery fields after removing the confounding effects of demographics with DemoVAE (see Figure~\ref{fig:zcorr}), not all correlations seem to be based on demographic confounds. Among the fields that remained significantly correlated were antipsychotics medication use and four PANSS symptom severity fields in the BSNIP dataset. In fact, \textit{Sendi et al.}, among others \cite{Bernard2017-ks}, reported changes in FC correlate with schizophrenia symptoms \cite{Sendi2021-vl}. Additionally, \textit{Chopra et al.} identified differential FC in schizophrenia patients taking antipsychotic medication compared to antipsychotic-naive patients \cite{Chopra2021-ep}. We believe the fact that DemoVAE quickly identifies clinical outcomes that are measurable by FC and unconfounded with respect to demographic information makes it a worthwhile contribution to the neuroimaging communities.

\section{Conclusion}
\label{sec:conclusion}

This paper proposes a new way to condition the well-known VAE model on demographic information by decorrelating demographic information during VAE training and incorporating this information into the decoder stage. This method of conditioning and training creates synthetic samples that recapitulate both group differences as well as individual subject variation in FC. We show that DemoVAE outperforms a traditional VAE in capturing the whole distribution of fMRI data. It is shown that most clinical questionnaire and computerized battery fields that are correlated with fMRI features are in fact confounded by the ability of fMRI features to predict demographics. By contrast, our DemoVAE model shows that several clinical outcomes related to schizophrenia are independent of demographic features. We hope this finding can shed lights on the appropriate future use of demographic information in neuroimaging. 

%We present this model in the hope that the insights gleened in this paper may be useful to the neuroimaging and machine learning communities.

\bibliographystyle{IEEEtran}
\bibliography{IEEEabrv,demovae2}

\end{document}